\documentstyle[bo99,epsfig]{article}

\title{Comptonization in X-ray bright neutron star globular cluster systems }
\author{M.Guainazzi$^1$, A.N.Parmar$^2$, T.Oosterbroek$^2$}
\affil{1) XMM SOC, VILSPA, ESA, Apartado 50727, E-28080 Madrid, Spain; 2) Astrophysics Division, Space Science Department of ESA, Postbus 299, NL-2200 AG Noordwijk, The Netherlands}

\begin{document}

\maketitle

\begin{abstract}
A BeppoSAX survey of bright Neutron Star (NS) systems in globular clusters, together with the results obtained on Galactic LMXRB, suggests that a two-component model is a reasonable description of the 0.1--200~keV spectra. At energies lower than a few keV, a thermal component dominates, either due to the direct view of the NS surface, to the emission of the boundary layer, or to a multi-temperature blackbody originating in the accretion disk. At higher energies, a power-law with an high-energy cutoff is the likely signature of thermal Comptonization of soft X-ray photons, probably produced by the same thermal mechanisms as above. The plasma optical depth (electron temperature) correlates (anticorrelates) with the X-ray luminosity, suggesting that the properties of the Comptonizing plasma are driven by the X-ray energy output.

\end{abstract}

BeppoSAX (Boella et al. 1997) is carrying out a systematic survey of X-ray bright globular cluster sources. More than 90\% of them (11 out of 12) are binary Neutron Star (NS) systems. The sources discussed in this paper are listed in Tab.1.

The "burster box" (see Fig.1)
was invented by Barret et al. (1996) to discriminate
%--------------------------  figure 1
\begin{figure}[h]
\centerline{\psfig{file=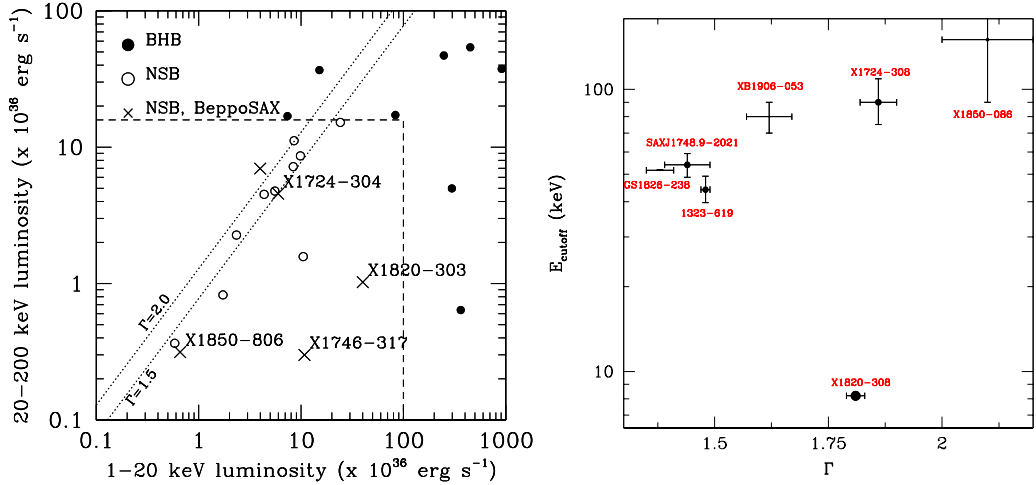, width=7cm}
\psfig{file=fig2.cps, width=7cm}
}
\caption[]{{\it Left panel}: 1--20~keV versus 20--200~keV luminosity for black hole ({\it filled circles}) and neutron star ({\it empty circles}) binary systems. The globular cluster sources observed by BeppoSAX are plotted with {\it crosses};
{\it Right panel}: cutoff energies versus spectral indices when a model constituted by a disk blackbody and a cut-offed power-law is applied to the LMXRB observed by BeppoSAX. All measures are from our analysis, except:  XB1916-053 (Church et al. 1999), 4U1323-69 (Ba\l uci\'nska-Church M. et al. 1999), GS1826-238 (in't Zand et al. 1999b), SAXJ1748.9-2021 (in `t Zand et al. 1999a) }
\label{fig1}
\end{figure}
%---------------------------------
between NS and Black Hole binaries (BHB). With the BeppoSAX observations (crosses) we can for the first time fill this box with simultaneous measures in the 1-20 and 20-200 keV energy bands (see also Barret et al. 1999). Most of the NS sources align within a strip corresponding to simple power-laws with photon index ($\Gamma$) between 1.5 and 2.0. However, a few exceptions exist. Is it possible to encompass also these exceptions in the same physical framework?

In several NS binaries, BeppoSAX measured high-energy cutoffs (see Fig.2 for
two illustrative cases).
Although there is no clear trend with the X-ray luminosity (proportional to the data point size in Fig.1), it is worthwhile to notice that the only source with ${\rm E_{cutoff} < 10}$~keV is that with ${\rm \log(L_X) \sim 38}$ (X1820-308).

In our sample, a model consisting of a disk blackbody and thermal Comptonization in a spherical geometry provides a good fit of the broadband BeppoSAX spectrum in all cases (see Fig.2). Some
%--------------------------  figure 3
\begin{figure}
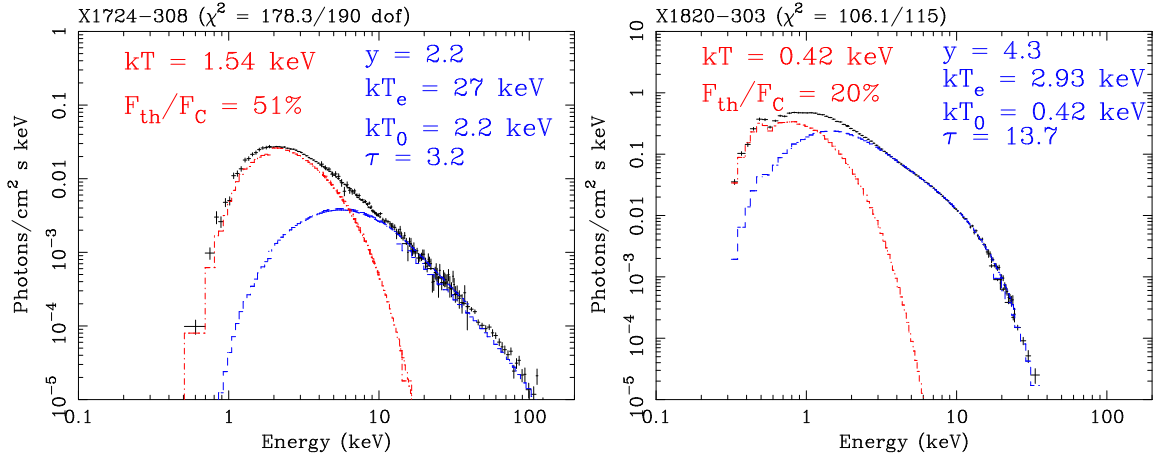

\centerline{\psfig{file=fig3a.cps, width=6cm, angle=-90}
\psfig{file=fig3c.cps, width=6cm, angle=-90}
}
\caption[]{X-ray spectral energy distributions as measured by BeppoSAX ({\it crosses}) compared with the best fit models for X1724-308 (${\rm L_{0.1-200 \ keV} \sim 10^{37}}$~erg~s$^{-1}$, {\it left}), and X1820-303 (${\rm L_{0.1-200 \ keV} \simeq 8 \times 10^{37}}$~erg~s$^{-1}$, {\it right}). In both cases, the model is constituted by Comptonization in spherical geometry ({\it high energy component}) and a disk blackbody component ({\it low energy component)}.}
\label{fig3}
\end{figure}
%---------------------------------
best-fit parameters are listed in Tab.1. The underlying scenario is a distribution of soft photons, originating in the accretion disk,
%--------------------------  Table 1
\begin{table}
\begin{footnotesize}
\begin{tabular}{lccccccc} \hline \hline
& ${\rm kT_e}$ & ${\rm \tau}$ & ${\rm kT_0}$ & ${\rm kT_{disk}}$ & ${\rm F_{disk}/F_{Compt}}$ & ${\rm \chi^2_{\nu}}$ \\
& & (keV) & (keV) & (keV) & & \\ \hline
X1724-308 (Terzan~2) & $27\pm^{11}_4$ & $3.2 \pm^{0.6}_{0.8}$ & $2.2 \pm^{0.3}_{0.2}$ & $1.54 \pm^{0.11}_{0.08}$ & 0.5 & 0.94 \\
X1746-376 (NGC6441) & $<110$ & $<3.7$ & $0.43 \pm 0.03$ & $2.79 \pm 0.04$ & 6.9 & 1.08 \\
X1820-303 (NGC6624) & $2.93 \pm 0.06$ & $13.7 \pm ^{0.5}_{0.4}$ & $0.46 \pm^{0.08}_{0.06}$ & $0.42 \pm 0.06$ & 0.2 & 0.92 \\
X1850-086 (NGC6712) & $110 \pm^{240}_{50}$ & $0.7 \pm^{0.2}_{0.4}$ & $0.15 \pm^{0.15}_{0.12}$ & $0.64 \pm ^{0.06}_{0.04}$ & 0.3 & 1.02 \\ \hline \hline
\end{tabular}
\end{footnotesize}
\caption{Best-fit parameters when the two components (disk blackbody +Comptonization) model is applied to the BeppoSAX observations of X-ray bright globular cluster (in brackets) sources. From left to right: Comptonizing plasma electron temperature and optical depth, input Wien distribution temperature, disk blackbody temperature, ratio between the disk blackbody and Comptonized fluxes, reduced chi-squared}
\label{tab1}
\end{table}
%--------------------------  Table 1
which is Comptonized in a region closer to the NS surface, {\it e.g.} the boundary layer. A model where the thermal emission originates from a single-temperature blackbody ({\it e.g.} the surface of the NS), and
the Comptonization occurs in a disk corona above the disk yields comparably good ${\rm \chi^2_{\nu}}$, and is discussed in Guainazzi et al. (1998). The properties of the Comptonization do not substantially differ in the two scenarios.
The ratio of the thermal versus the Comptonization X-ray fluxes is 0.2--0.5. The ratio of the thermal component versus input Comptonization temperatures is on the average ~0.7 (if one does not consider X1746-371, see below), and is consistent with the value of 1 in two out of three cases within the statistical uncertainties. 
In X1746-371 (Parmar et al. 1999), 
the Comptonization is strongly absorbed. This is a "dipping" source, thus likely to be observed at high inclination. This supports the idea that the Comptonization occurs in a region smaller than that, where the thermal component originates
[see, however, a different point of view in Church et al. (1999)].

The Comptonizing plasma electron temperature and optical depth correlate cleanly with luminosity (see Fig.~3).
%--------------------------  figure 4
\begin{figure}[h]
\centerline{\psfig{file=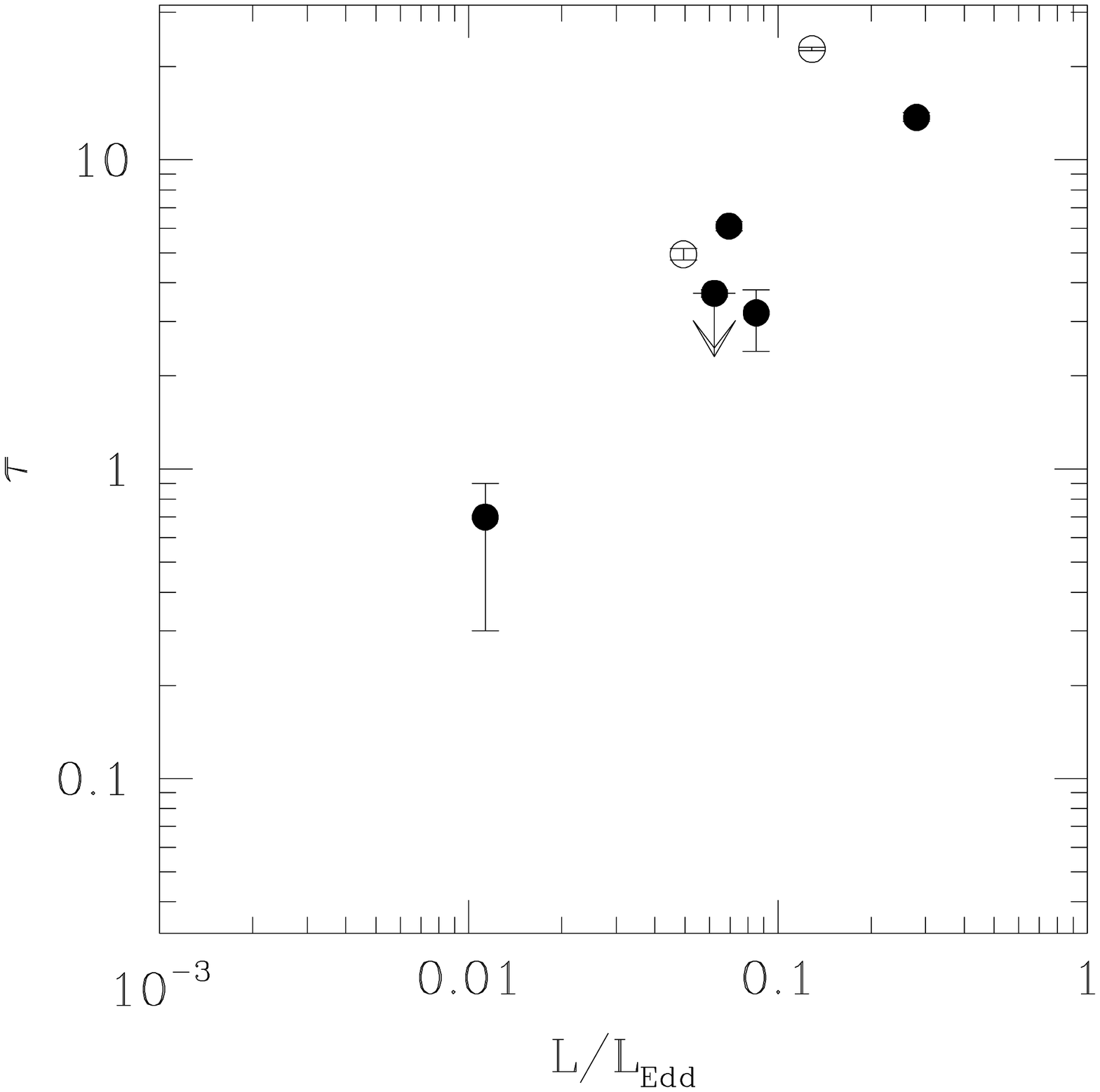, width=7cm}
\psfig{file=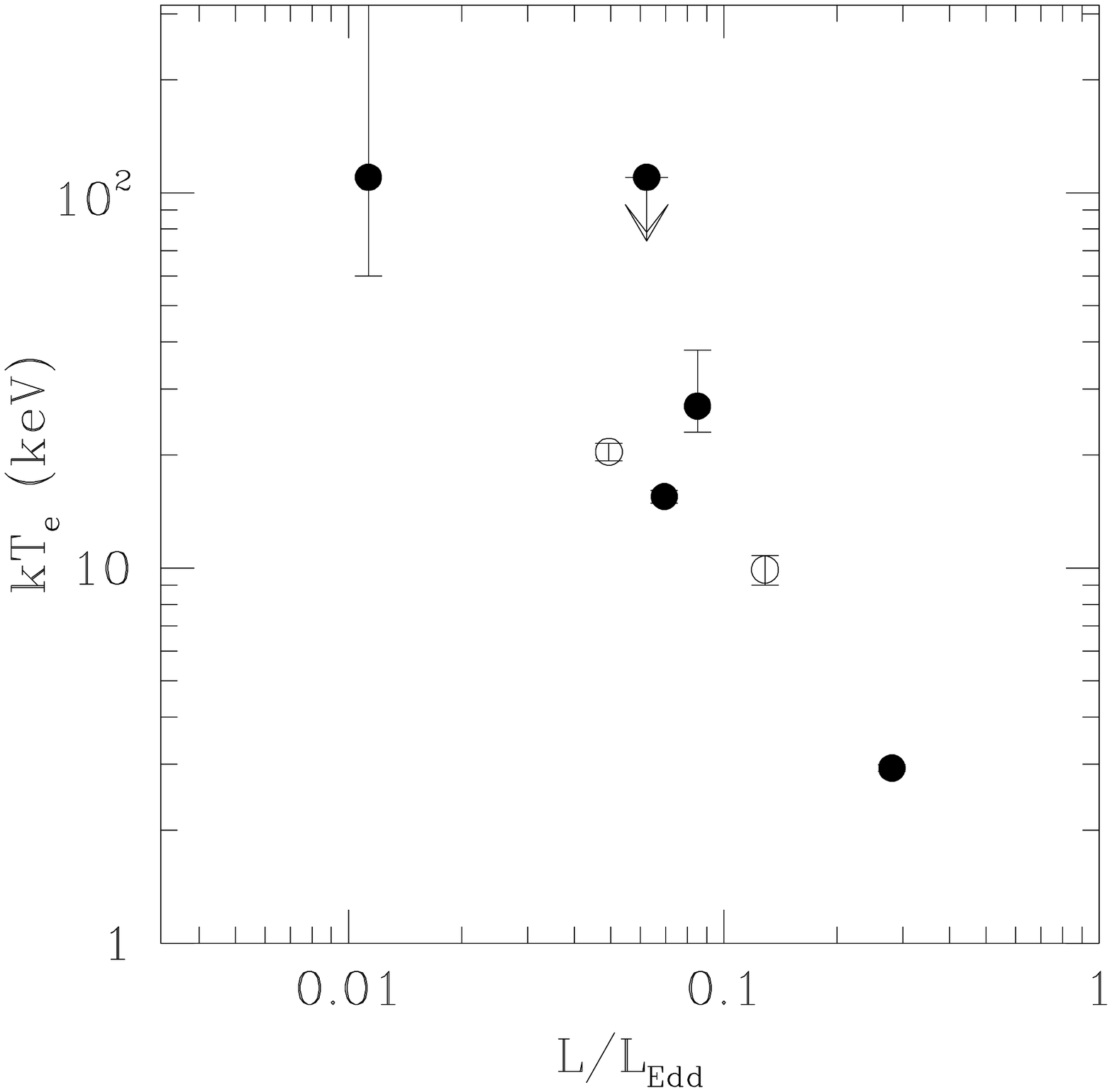, width=7cm}
}
\caption[]{Optical depth ({\it left panel}) and Comptonizing plasma electron temperature ({\it right panel}) versus luminosity correlation in our sample, plus SAXJ1748.9-2021 (in' t Zand et al. 1999a; {\it filled circles}. The Galactic LMXRB X1822-371 (Parmar et al. 2000) and GC1826-238 (in't Zand et al. 1999b) are also displayed with {\it empty circles}}
\label{fig6}
\end{figure}
%--------------------------------
In the above scenario (thermal photons from the disk; Comptonization in the boundary layer), the correlation may be qualitatively explained. If the X-ray luminosity of the boundary layer is proportional to the accretion rate (King 1995), the higher ${\rm \dot{m}}$, the higher is the plasma influx beyond the boundary layer radius. Hence the Comptonizing plasma optical depth is higher and the Compton cooling is more efficient, yielding a lower Comptonizing electron temperature.

\end{document}